\documentclass[12pt]{article}
  \usepackage{amsfonts}
  \usepackage{amsmath}
\usepackage{amssymb}
\usepackage{amscd}
\usepackage[dvips]{graphicx}
\begin{document}
~~
\bigskip
\bigskip
\bigskip
\bigskip
\bigskip
\begin{center}
\section*{{%\large\bf
QUANTIZATION OF $\kappa$-DEFORMED FREE FIELDS AND $\kappa$-DEFORMED
OSCILLATORS}}
\end{center}
\bigskip
\bigskip
\bigskip
\begin{center}
{{\large\bf ${\rm Marcin\;Daszkiewicz}^{1}$, ${\rm Jerzy\;
Lukierski}^{1,2}$, ${\rm Mariusz\;Woronowicz}^{1}$}}
\end{center}
\begin{center}
\bigskip

{%\large
${\rm{~^{1}Institute\; of\; Theoretical\; Physics}}$}

{%\large
${\rm{ University\; of\; Wroc{\l}aw\; pl.\; Maxa\; Borna\; 9,\;
50-206\; Wroc{\l}aw,\; Poland}}$}

{%\large
${\rm{ e-mail:\; marcin,lukier,woronow@ift.uni.wroc.pl}}$}
\bigskip

{%\large
${\rm{~^{2}Department\; of\; Theoretical\; Physics}}$}

{%\large
${\rm{ University\; of\; Valencia,\; 46100\; Burjassot\;
(Valencia),\; Spain}}$}

{%\large
${\rm{ e-mail:\; lukier@ift.uni.wroc.pl}}$}

\end{center}
%\bigskip
%\bigskip
%\bigskip
%\bigskip
%\bigskip
\bigskip
\begin{abstract}
We describe the deformed E.T. quantization rules for
$\kappa$-deformed free quantum fields, and relate these rules with
the $\kappa$-deformed algebra of field oscillators.
\end{abstract}
\bigskip
%\bigskip
%\bigskip
%\bigskip

\section{Introduction.}

Recently, there were considered in several papers \cite{1}-\cite{7}
the field quantization rules for canonically deformed QFT, with the
noncommutative space-time coordinates satisfying the relation
\begin{equation}
[\;{\hat x}_{\mu},{\hat x}_{\nu}\;] =
i\theta_{\mu\nu}\;\;\;\;\;\;\;\;\;\;(\theta_{\mu\nu} = {\rm
const})\;. \label{canonical}
\end{equation}
The deformed QFT can be represented by classical fields if we use
the Weyl map $f({\hat{x}}) \to f(x)$ provided that the product of
commutative fields at different space-time points $x$, $y$ is given
by a bilocal Moyal star product
\begin{equation}
f(x)\star_{\theta}g(y) = {\rm
exp}\left[\frac{i}{2}\theta^{\mu\nu}{\partial}_\mu^{(x)}
{\partial}_\nu^{(y)} \right]f(x)g(y)\;. \label{biproduct}
\end{equation}
The choice of (\ref{biproduct}) corresponds to the following
extension of the relation (\ref{canonical}) for a pair ${\hat
x}_{\mu}$, ${\hat y}_{\mu}$ of noncommutative space-time coordinates
\begin{equation}
[\;{\hat x}_{\mu},{\hat y}_{\nu}\;] = [\;{\hat y}_{\mu},{\hat
y}_{\nu}\;] = i\theta_{\mu\nu}\;. \label{canonical2}
\end{equation}
The free $\theta^{\mu\nu}$-deformed KG field is described on
commutative Minkowski space by standard free action\footnote{One can
show that $\int d^4xf(x)\star_{\theta}f(x) = \int d^4xf(x)f(x)$;
other argument is that the $\theta^{\mu\nu}$-deformation is
generated by twisted Poincare algebra \cite{2}-\cite{7}, with
unmodified mass and spin square Casimirs.}
\begin{equation}
{\cal{L}}_0 = \frac{1}{2} \left[ (\partial_{\mu}\varphi_{0}(x))^2 -
m^2 \varphi_{0}(x)^2 \right]\;, \label{action}
\end{equation}
and we quantize the fields through a deformed E.T. commutation
relations
\begin{equation}
[\; \varphi_{0}(t,\vec{x}), \pi_{0}(t,\vec{y})\;]_{\star_{\theta}} =
\varphi_{0}(t,\vec{x})\star_{\theta}\pi_{0}(t,\vec{y}) -
\pi_{0}(t,\vec{y})\star_{\theta}\varphi_{0}(t,\vec{x}) =
i\delta^{(3)}(\vec{x}-\vec{y}) \;, \label{fieldCCR}
\end{equation}
where $\pi_0(x) = \frac{\partial
{\cal{L}}_0}{\partial(\partial_{t}\varphi_{0})} = \partial_t
\varphi_{0}(x)$, and
\begin{equation}
[\; \varphi_{0}(t,\vec{x}),
\varphi_{0}(t,\vec{y})\;]_{\star_{\theta}} = [\; \pi_{0}(t,\vec{x}),
\pi_{0}(t,\vec{y})\;]_{\star_{\theta}} = 0\;. \label{fieldCCR1}
\end{equation}

Let us introduce the Fourier transform defining creation and
annihilation operators as follows
\begin{eqnarray}
{\hat \varphi}_{0}({\hat x}) &=& \frac{1}{(2\pi)^{3/2}} \int
{d^4{p}}~A(p){\rm e}^{ ip{\hat x}} ~\delta (p^2 - m^2)\nonumber \\
&&~~\vspace{1.0cm} \nonumber \\
&=& \left.\frac{1}{(2\pi)^{3/2}} \int \frac{d^3\vec{p}}
{2\omega(\vec{p})}\; \left (a(\vec{p})\; {\rm e}^{ ip{\hat x}}  +
a^\dag(\vec{p})\; {\rm e}^{-ip{\hat x}} \right)\right|_{p_0 =
\omega(\vec{p})}\;, \label{fourier}
\end{eqnarray}
where $\omega(\vec{p}) = \sqrt{\vec{p}^2 + m^2}$ and $a(\vec{p}) =
A(\omega(\vec{p}),\vec{p})$, $a^{\dag}(\vec{p}) =
A(-\omega(\vec{p}),-\vec{p})$. In order to obtain the relations
(\ref{fieldCCR}), (\ref{fieldCCR1}) one should postulate that the
oscillators $a(\vec{p})$, $a^\dag(\vec{p})$ satisfy the following
deformed algebra
\begin{eqnarray}
&&[\;a^\dag(\vec{p}),a(\vec{q})\;]_{\circ_{\theta}}  =
2\omega(\vec{p})\delta^{(3)}(\vec{p}-\vec{q})\;, \label{aa} \\
&&~~\vspace{1.0cm} \nonumber \\
&&[\;a(\vec{p}),a(\vec{q})\;]_{\circ_{\theta}}  =
[\;a^\dag(\vec{p}),a^\dag(\vec{q})\;]_{\circ_{\theta}}=0 \;,
\label{aa1}
\end{eqnarray}
with multiplication $\circ_{\theta}$ defined by
\begin{eqnarray}
&&a(\vec{p}) \circ_{\theta}a(\vec{q}) = {\rm e}^{-\frac{i}{2}p\theta
q}a(\vec{q}) a(\vec{p})\;,\label{m}\\
&&a^\dag(\vec{p}) \circ_{\theta}a^\dag(\vec{q}) = {\rm e}^{-\frac{i}{2}p\theta
q}a^\dag(\vec{q}) a^\dag(\vec{p})\;,\\
&&a^\dag(\vec{p}) \circ_{\theta}a(\vec{q}) = {\rm e}^{\frac{i}{2}p\theta
q}a(\vec{q}) a^\dag(\vec{p})\;,\\
&&a(\vec{p}) \circ_{\theta}a^\dag(\vec{q}) = {\rm e}^{\frac{i}{2}p\theta
q}a^\dag(\vec{q}) a(\vec{p})\;,
\end{eqnarray}
Using
\begin{equation}
e^{ip{x}}\star_{\theta}e^{iq{y}}=e^{i(p{x}+q{y})-\frac{i}{2}p\theta
q}\;, \label{exptwe}
\end{equation}
one gets that
\begin{eqnarray}
&&\varphi(x)\star_{\theta}\varphi(y)=\frac{1}{(2\pi)^3}\int d^4p\int
d^4q \;\delta(p^2-m^2)\delta(q^2-m^2)A(p)A(q)\;e^{ipx}
\star_\theta e^{iqy}\qquad\label{efe}\\
&&~~~~~~~~~~~~~~~=\frac{1}{(2\pi)^3}\int d^4p\int  d^4q \;\delta(p^2-m^2)\delta(q^2-m^2)A(p)\circ_\theta A(q)\;
e^{i(px+qy)}\nonumber\\
&&~~~~~~~~~~~~~~:=\varphi(x)\circ_\theta \varphi(y)\;,\nonumber
\end{eqnarray}
and from the relations (\ref{aa}), (\ref{aa1}) follows that
\begin{equation}
[\; \varphi_{0}({x}), \varphi_{0}({y})\;]_{\circ_{\theta}} =
i\Delta(x-y;m^2)\;, \label{pauli}
\end{equation}
where on rhs we obtain the fourdimensional classical Pauli-Jordan
commutator function. We see that in field-theoretic framework  the
noncommutativity of space-time coordinates described by the star
product (\ref{biproduct}) we express  equivalently by the
deformation (\ref{aa})-(\ref{exptwe}) of the oscillator algebra.

In this note we would like to present similar construction
describing the quantization of $\kappa$-deformed free field
\cite{11},\cite{12}. We would like however to point out that:

i) The $\kappa$-deformation modifies the free action of the scalar
field; the deformed KG operator will be given by the
$\kappa$-deformed mass Casimir,

ii) The fourmomentum coproduct for $\kappa$-Poincare algebra is
deformed. Keeping this nonsymmetric addition law it is desirable
however to introduce the particle states with classical addition law
of their fourmomenta,

iii) The quantization rules should lead to the quantum field
satisfying some sort of deformed $\kappa$-causality condition. The
$\kappa$-deformations do not preserve the notion of standard light
cone. We shall derive that the quantization rules for free
noncommutative scalar field should reproduce the
$\kappa$-deformation of the formula (\ref{pauli})
\begin{equation}
\Delta(x;m^2) \to \Delta_{\kappa}(x;m^2) = \frac{i}{(2\pi)^3} \int
d^4p ~\epsilon(p_0)~ \delta \left(C_{2}^{\kappa}(p)
 - m^2 \right) {\rm e}^{i{p}{x}}\;, \label{kappapauli}
\end{equation}
where $C_{2}^{\kappa}(p)$ describes the $\kappa$-deformed mass
Casimir of $\kappa$-Poincare Hopf algebra, and
$\Delta_{\kappa}(x;m^2)$ is an example of $\kappa$-causal function.

\section{Free field on $\kappa$-deformed space-time, quantization rules
and $\kappa$-deformed oscillators algebra.}

In this note we shall use modified, so-called symmetric
bicrossproduct basis \cite{8}-\cite{9} of $\kappa$-Poincare algebra
\cite{13p}-\cite{15}, with standard $\kappa$-deformed mass-shell
\cite{13}
\begin{equation}
C_{\kappa}^2 (\vec{P},P_0) = \left (2\kappa\sinh
\left(\frac{P_0}{2\kappa}\right)\right)^2 - \vec{P}^2\;,
\label{casimir}
\end{equation}
providing classical formulae for the coinverse $(S(I_i) = -I_i~;~I_i
= P_{\mu}, M_{\mu\nu})$. The Abelian Hopf subalgebra of fourmomenta
$([\,P_{\mu},P_{\nu}\,] = 0)$ is endowed with the coproduct
\begin{equation}
\Delta(P_0) = P_0\otimes 1 + 1\otimes P_0\;\;\;,\;\;\;\Delta(P_i)
=P_i\otimes {\rm e}^{\frac{P_0}{2\kappa}} + {\rm
e}^{-\frac{P_0}{2\kappa}}\otimes P_i\;, \label{copro}
\end{equation}
and dual Hopf subalgebra, describing $\kappa$-deformed Minkowski
space-time \cite{16},\cite{15},\cite{17}, has a form
\begin{equation}
[\;{\hat x}_{0},{\hat x}_{i}\;] = \frac{i}{\kappa}{\hat
x}_{i}\;\;\;,\;\;\; [\;{\hat x}_{i},{\hat x}_{j}\;] =
0\;\;\;;\;\;\;\Delta({\hat x}_{\mu}) = {\hat x}_{\mu}\otimes 1 +
1\otimes {\hat x}_{\mu}\;.  \label{minkowski}
\end{equation}

The $\kappa$-deformed free quantum scalar field on noncommutative
Minkowski space (\ref{minkowski}) is given by the following
noncommutative Fourier transform
\begin{eqnarray}
{\hat \phi} ({\hat x}) &=& \frac{1}{(2\pi)^{\frac{3}{2}}} \int
d^4p\; A(p_0,\vec{p})\;\delta \left(C^2_{\kappa} (\vec{p},p_0) -
M^2\right)\vdots {\rm e}^{ip_\mu \hat{x}^\mu}\vdots =
~~~~~~~\nonumber \\
&&~~\vspace{1.0cm} \nonumber \\
&=& \left.\frac{1}{(2\pi)^{\frac{3}{2}}} \int
\frac{d^3\vec{p}}{2\Omega_{\kappa}(\vec{p})}\;\left( a( \vec{p})\;
\vdots {\rm e}^{ip_\mu \hat{x}^\mu}\vdots + a^{\dag}( \vec{p})\;
\vdots {\rm e}^{-ip_\mu \hat{x}^\mu}\vdots \right)\right|_{p_0 =
\omega_{\kappa}(\vec{p})}\;, \label{field}
\end{eqnarray}
where we have chosen
\begin{equation}
\vdots {\rm e}^{ip_\mu \hat{x}^\mu}\vdots ={\rm
e}^{\frac{i}{2}p_0{\hat x}^{0}}{\rm e}^{ip_i{\hat x}^{i}} {\rm
e}^{\frac{i}{2}p_0{\hat x}^{0}} \;, \label{exp}
\end{equation}
and performing mass-shell Dirac delta we get
\begin{equation}
\omega_\kappa(\vec{p}) = 2\kappa~{\rm arcsinh}\left
(\frac{\sqrt{\vec{p}^2 +M^2}}{2\kappa}
\right)\;\;\;,\;\;\;\Omega_{\kappa}(\vec{p}) = \kappa
\sinh\left(\frac{\omega_{\kappa}(\vec{p})}{\kappa}\right)\;,
\label{omega}
\end{equation}
\begin{equation}
a_{\kappa}(\vec{p}) =
A(\omega_{\kappa}(\vec{p}),\vec{p})\;\;\;,\;\;\; a_{\kappa}^\dag
(\vec{p}) = A(-\omega_{\kappa}(\vec{p}),-\vec{p})\;.
\label{creation}
\end{equation}
In order to represent the algebra of noncommutative fields
(\ref{field}) by the classical ones, on commutative Minkowski space,
one should introduce the homomorphic $\star_{\kappa}$-multiplication
on classical functions obtained by the Weyl map $\vdots {\rm
e}^{ip_\mu\hat{x}^\mu}\vdots \rightarrow {\rm e}^{ip_\mu{x}^\mu}$.
Because
\begin{equation}
\vdots {\rm e}^{ip_\mu\hat{x}^\mu}\vdots \cdot \vdots{\rm
e}^{iq_\mu\hat{x}^\mu}\vdots = \vdots{\rm
e}^{i\Delta_\mu(p,q)\hat{x}^\mu}\vdots\;, \label{point1}
\end{equation}
with (see (\ref{copro}))
\begin{equation}
\Delta_\mu(p,q) = \left(\Delta_0 = p_0 + q_0,~ \Delta_i = p_i{\rm
e}^{\frac{q_0}{2\kappa}} + q_i{\rm
e}^{-\frac{p_0}{2\kappa}}\right)\;, \label{ddelta}
\end{equation}
the $\star_{\kappa}$-product reproducing multiplication
(\ref{point1}) has a form
\begin{equation}
{\rm e}^{ip_\mu{x}^\mu} \star_{\kappa} {\rm e}^{iq_\mu{y}^\mu} =
{\rm e}^{i(p_0x^0+q_0y^0) + \Delta_i(\vec{p},\vec{q})x^i }\;.
\label{onepoint}
\end{equation}
We extend the multiplication (\ref{onepoint}) to bilocal products of
functions as follows
\begin{equation}
{\rm e}^{ip_\mu{x}^\mu} \star_{\kappa} {\rm e}^{iq_\mu{y}^\mu} =
{\rm e}^{i(p_0x^0+q_0y^0) + (p_i{\rm e}^{\frac{q_0}{2\kappa}}x^i +
q_i{\rm e}^{-\frac{p_0}{2\kappa}}y^i)}\;, \label{twopoint}
\end{equation}
where the formula (\ref{twopoint}) corresponds to the following
cross relations of two $\kappa$-Minkowski coordinates
\cite{11},\cite{12}
\begin{eqnarray}
&&[\,x_0,y_i\,] = \frac{i}{\kappa}y_i\;\;\;,\;\;\;\;[\,x_0,y_0\,] =
0\;, \label{ttwo} \\
&&[\,x_i,y_i\,] = 0\;\;\;,\;\;\;\;[\,x_i,y_0\,] =
-\frac{i}{\kappa}x_i\;. \nonumber
\end{eqnarray}
We introduce the $\kappa$-star product of two free classical fields
in the following way
$$
\varphi(x)\star_{\kappa} \varphi(y) = \frac{1}{(2\pi)^{3}} \int
d^4p\;\int d^4q\; A(p_0,\vec{p})A(q_0,\vec{q})\cdot{\rm
e}^{i(p_0x^0+q_0y^0) + (p_i{\rm e}^{\frac{q_0}{2\kappa}}x^i +
q_i{\rm e}^{-\frac{p_0}{2\kappa}}y^i)}   \nonumber
$$

\begin{equation}
~~~~~~~~~~~~~~~~~ \delta ( C^2_{\kappa} (\vec{p}{\rm
e}^{\frac{q_0}{2\kappa}},p_0) - M^2)~ \delta ( C^2_{\kappa}
(\vec{q}{\rm e}^{-\frac{p_0}{2\kappa}},q_0) - M^2 )\;.
~~\vspace{0.3cm} \label{novum}
\end{equation}
In the definition (\ref{novum}) besides the use of formula
(\ref{twopoint}) we also additionally shift the value of
threemomentum variable in the mass-shell conditions, in accordance with
the threemomentum values obtained in the coproduct (\ref{ddelta}). 
One can also observe that the action of the fourmomentum generator on the products of oscillators and exponentials in (\ref{novum}) has the form
\begin{equation}
P_\mu\triangleright[A(p)A(q)]=\chi_\mu[A(p)A(q)]\,,\qquad P_\mu\triangleright[e^{ipx}\star_\kappa e^{iqy}]=\chi_\mu[e^{ipx}\star_\kappa e^{iqy}]\,,
\label{newprodd}
\end{equation}
and it is described by the coproduct (\ref{copro}) $(P_\mu\triangleright[fg]=\omega\circ\Delta(P_\mu)(f\otimes g)$ and $ \omega(f\otimes g)=fg)$. 
One gets that the eigenvalues in both relations (\ref{newprodd}) are the same 
\begin{equation}
\chi_0=p_0+q_0\,,\qquad\chi_i=p_i e^{q_0/2\kappa}+q_i e^{-p_0/2\kappa}\,,
\end{equation}
where $P_\mu\triangleright A(p)=p_\mu A(p)$ and $P_\mu\triangleright e^{ipx}=p_\mu e^{ipx}$.

The multiplication (\ref{novum}) of the $\kappa$-deformed free fields after the
change of variables
\begin{equation}
{\cal{P}}_{0} = p_{0}\;\;\;,\;\;\;{\cal{P}}_{i} = p_{i}{\rm
e}^{\frac{q_0}{2\kappa}}\;\;\;,\;\;\;{\cal{Q}}_{0} =
q_{0}\;\;\;,\;\;\;{\cal{Q}}_{i} = q_{i}{\rm
e}^{-\frac{p_0}{2\kappa}}\;. \label{variables}
\end{equation}
leads to the formula
\begin{eqnarray}
 &&\varphi(x)\star_{\kappa} \varphi(y)  = \frac{1}{(2\pi)^{3}}
\int d^4{\cal P}\;\int d^4{\cal Q}\; {\rm e}^{\frac{3({\cal{P}}_0 -
{\cal{Q}}_{0})}{2\kappa}} A({\cal{P}}_0,{\rm e}^{-\frac{{\cal
Q}_0}{2\kappa}}{{\cal{P}}}_i) A({\cal{Q}}_0,{\rm e}^{\frac{{\cal
P}_0}{2\kappa}}{{\cal{Q}}}_i)
\nonumber \\
&&~~ \nonumber \\
&&~~~~~~\cdot{\rm e}^{i({\cal P}_0x^0+{\cal Q}_0y^0) + ({\cal
P}_ix^i + {\cal Q}_iy^i)} \delta ( C^2_{\kappa} (\vec{{\cal
P}},{\cal P}_0) - M^2)~ \delta ( C^2_{\kappa} (\vec{{\cal Q}},{\cal
Q}_0) - M^2 )\;.\label{newstar1}\\
&&~~\vspace{0.3cm} \nonumber
\end{eqnarray}
We see that if one introduces the following $\kappa$-deformed
multiplication for creation/anihilation operators (\ref{creation})
$(F_\kappa^{(2)}({\cal{P}}_0,{\cal{Q}}_0) ={\rm
e}^{{{\frac{3}{2\kappa}({\cal{P}}_0 - {\cal{Q}}_0)}}})$
\begin{eqnarray}
a_\kappa ({{\cal{P}}})\circ_{\kappa}
a_\kappa({{\cal{Q}}})&=&F_\kappa^{(2)}({\cal{P}}_0,{\cal{Q}}_0)\;a_\kappa
\left({\cal{P}}_0,{\rm
e}^{-\frac{{\cal{Q}}_0}{2\kappa}}\vec{{\cal{P}}} \right) a_\kappa
\left({\cal{Q}}_0,{\rm
e}^{\frac{{\cal{P}}_0}{2\kappa}}\vec{{\cal{Q}}}
\right)\;,\label{multi0}\\
&&~~\vspace{1.0cm} \nonumber \\
a_\kappa^\dag ({\cal{P}})\circ_{\kappa} a_\kappa^\dag ({\cal{Q}})&=&
F_\kappa^{(2)}(-{\cal{P}}_0,-{\cal{Q}}_0)\;a_\kappa^\dag
\left({\cal{P}}_0,{\rm
e}^{\frac{{\cal{Q}}_0}{2\kappa}}\vec{{\cal{P}}} \right)
a_\kappa^\dag \left({\cal{Q}}_0,{\rm
e}^{-\frac{{\cal{P}}_0}{2\kappa}}\vec{{\cal{Q}}}
\right)\;,\label{multi1}\\
&&~~\vspace{1.0cm} \nonumber \\
a_\kappa^\dag ({\cal{P}})\circ_{\kappa} a_\kappa({\cal{Q}})&=&
F_\kappa^{(2)}(-{\cal{P}}_0,{\cal{Q}}_0)\;a_\kappa^\dag
\left({\cal{P}}_0,{\rm
e}^{-\frac{{\cal{Q}}_0}{2\kappa}}\vec{{\cal{P}}} \right) a_\kappa
\left({\cal{Q}}_0,{\rm
e}^{-\frac{{\cal{P}}_0}{2\kappa}}\vec{{\cal{Q}}}
\right)\;,\label{multi2}\\
&&~~\vspace{1.0cm} \nonumber \\
a_\kappa ({\cal{P}})\circ_{\kappa} a_\kappa^\dag({\cal{Q}})&=&
F_\kappa^{(2)}({\cal{P}}_0,-{\cal{Q}}_0)\;
a_\kappa\left({\cal{P}}_0,{\rm
e}^{\frac{{\cal{Q}}_0}{2\kappa}}\vec{{\cal{P}}} \right)
a_\kappa^\dag \left({\cal{Q}}_0,{\rm
e}^{\frac{{\cal{P}}_0}{2\kappa}}\vec{{\cal{Q}}}
\right)\;,\label{multi3}
\end{eqnarray}
by using the multiplication rules (\ref{multi0})-(\ref{multi3}) for
any field operators pair we get \cite{12}
\begin{equation}
\varphi(x)\star_{\kappa} \varphi(y) = \varphi(x)\circ_{\kappa}
\varphi(y)\;. \label{identity}
\end{equation}

Let us now define the $\kappa$-deformed canonical momentum as
follows
\begin{equation}
\pi(t,\vec{x}) =
\partial_t^{\kappa}\varphi(t,\vec{x})\;,\label{momentum}
\end{equation}
where
\begin{equation}
\partial_t^{\kappa} \equiv i\kappa \sinh\left (\frac{1}{i}\frac{\partial_t}{\kappa}\right)
= \kappa \sin \left
(\frac{\partial_t}{\kappa}\right)\;\;\;{~}_{\overrightarrow{{\;\;\;\kappa
\rightarrow \infty}\;\;\;}}\;\;\;
\partial_t\;,
\label{qder}
\end{equation}
denotes the $\kappa$-deformed "quantum" derivative in time
direction\footnote{In fact the quantum time derivative (\ref{qder}) in bicovariant differential calculus for $\kappa$-deformed Poincare symmetries \cite{koma} has additional term $(\partial^\kappa_t=\kappa\sin\left(\frac{\partial_t}{\kappa}\right)+i\frac{\Delta}{2\kappa})$ but the second term does not provide any contribution to the equal time limit (\ref{kappafieldCCR}).}. We perform quantization of the $\kappa$-deformed fields
through the following deformed E.T. commutation relations
\begin{equation}
[\; \varphi(t,\vec{x}), \pi(t,\vec{y})\;]_{\star_{\kappa}} =
\varphi(t,\vec{x})\star_{\kappa}\pi(t,\vec{y}) -
\pi(t,\vec{y})\star_{\kappa}\varphi(t,\vec{x}) =
i\delta^{(3)}(\vec{x}-\vec{y}) \;, \label{kappafieldCCR}
\end{equation}
and
\begin{equation}
[\; \varphi(t,\vec{x}), \varphi(t,\vec{y})\;]_{\star_{\kappa}} = [\;
\pi(t,\vec{x}), \pi(t,\vec{y})\;]_{\star_{\kappa}} = 0\;.
\label{kappafieldCCR1}
\end{equation}
From the relations (\ref{kappafieldCCR}) and
(\ref{kappafieldCCR1}) one gets the field
oscillators (\ref{creation}) satisfying the following $\kappa$-deformed
algebra $([\;A,B\;]_{\circ_{\kappa}} = A\circ_{\kappa}B -
B\circ_{\kappa}A)$
\begin{eqnarray}
&&[\;a^\dag(\vec{p}),a(\vec{q})\;]_{\circ_{\kappa}}  =
2\Omega_{\kappa}(\vec{p})~\delta^{(3)}(\vec{p}-\vec{q})\;,\label{aaa0}\\
&&~~\vspace{1.0cm} \nonumber \\
&&[\;a(\vec{p}),a(\vec{q})\;]_{\circ_{\kappa}}  =
[\;a^\dag(\vec{p}),a^\dag(\vec{q})\;]_{\circ_{\kappa}} = 0 \;.
\label{aaa}
\end{eqnarray}
In fact, using the formula (\ref{identity}) and CCR relations
(\ref{aaa0}), (\ref{aaa}) one derives the fourdimensional
$\kappa$-deformed Pauli-Jordan commutator function (see
(\ref{kappapauli}))
\begin{equation}
[\; \varphi_{0}({x}), \varphi_{0}({y})\;]_{\star_{\kappa}} =
i\Delta_{\kappa}(x-y;m^2)\;,~~~~~~~~ \label{kappapauli1}
\end{equation}
with the following properties
\begin{equation}
\Delta_\kappa (x-y;M^2)|_{x_0 = y_0}= 0\;\;\;,\;\;\;
\partial_t^{\kappa}\Delta_\kappa (x-y;M^2)|_{x_0 = y_0}=
\delta^{(3)}(\vec{x}-\vec{y})\;. \label{delta}
\end{equation}
We see that similarly as in the case of
$\theta^{\mu\nu}$-deformation (see (\ref{fieldCCR}),
(\ref{fieldCCR1})), we recover E.T. relations (\ref{kappafieldCCR})
and (\ref{kappafieldCCR1}) for $\kappa$-deformed free field and its
canonical momentum (\ref{momentum}).

\section{$\kappa$-oscillators algebra and Abelian fourmomenta addition law.}

The definition of binary $\circ_{\kappa}$-product of
$\kappa$-oscillators can be extended to any monomial of creation and
annihilation operators \cite{11},\cite{12}. The n-fold product of
$\kappa$-deformed oscillators looks as follows\footnote{The formula
(\ref{n1}) if extended to negative energies describes also the
product of annihilation operators (see (\ref{creation})).}
\begin{equation}
a_\kappa({p}^{(1)} )\circ \cdots \circ a_\kappa({p}^{(n)} ) =
F_\kappa^{(n)}(p_0^{(1)},\ldots,p_0^{(n)})
a_\kappa\left(p_0^{(1)},\vec{{\cal P}}_{n}^{(1)} \right) \cdots
a_\kappa\left(p_0^{(n)},\vec{{\cal P}}_{n}^{(n)} \right)\;,
\label{n1}
\end{equation}
with
\begin{equation}
{\vec {\cal P}}_n^{(k)} =  \exp \frac{1}{2\kappa} \left(
{\sum\limits_{j=1}^{k-1}}p_0^{(j)} -
{\sum\limits_{j=k+1}^{n}}p_0^{(j)}\right)
\vec{{p}}^{~(k)}\;,\label{calmom}
\end{equation}
and
\begin{equation}
F_{\kappa}^{(n)}(p_0^{(1)},\ldots,p_0^{(n)}) =   {\rm
exp}\left({\frac{3}{2\kappa}{\sum\limits_{k=1}^{n}(n+1-2k)}p_0^{(k)}}\right)
\;.
\end{equation}

Let us introduce the vacuum in standard way
\begin{equation}
a_\kappa^\dag (\vec{p})|0> = 0\;\;\;,\;\;\; P_\mu\,|0> = 0\;,
\label{vacuum}
\end{equation}
where $P_\mu$ is the fourmomentum generator. The n-particle state we
define by
\begin{equation}
|\vec{p}^{~(1)},\ldots,\vec{p}^{~(k)},\ldots,\vec{p}^{~(n)}> =
a_\kappa(p^{(1)})\circ_{\kappa} \ldots \circ_{\kappa}
a_\kappa(p^{(k)})\circ_{\kappa} \ldots \circ_{\kappa}
a_\kappa(p^{(n)})|0> \label{state}\,.
\end{equation}
Let us consider in detail the two-particle state
\begin{equation}
|\vec{p},\vec{q}> =  a_\kappa(p_0,\vec{p})\circ_{\kappa}
a_\kappa(q_0,\vec{q})|0> = a_\kappa({\vec{\cal
P}}(\vec{p},q_0))\,a_\kappa({\vec{\cal Q}}(\vec{q},p_0))|0>\;,
\label{two}
\end{equation}
where
\begin{equation}
{\vec{\cal P}} = {\rm
e}^{-\frac{q_0}{2\kappa}}{\vec{p}}\;\;\;,\;\;\; {\vec{\cal Q}} =
{\rm e}^{\frac{p_0}{2\kappa}}{\vec{q}}\;.
 \label{ttt}
\end{equation}
Using the formula $({\cal Q}_0 = q_0,\;{\cal P}_0 = p_0)$
\begin{eqnarray}
 P_i\,\triangleright\left[\,a_\kappa({\vec{\cal P}})\,a_\kappa({\vec{\cal
Q}})\,\right] &=& \omega \left(\Delta(P_i)
(a_\kappa({\vec{\cal P}})\otimes a_\kappa({\vec{\cal Q}}))\right)
\nonumber \\
&&~~ \nonumber \\
 &=&\left( {\rm e}^{\frac{{\cal Q}_0}{2\kappa}}{\cal P}_i +
 {\rm e}^{-\frac{{\cal P}_0}{2\kappa}}{\cal Q}_i\right)
(a_\kappa({\vec{\cal P}})\, a_\kappa({\vec{\cal Q}}))
\label{aaction}\\
&&~~ \nonumber \\
 &=&(p_i+q_i)\,a_\kappa({\vec{\cal P}})\, a_\kappa({\vec{\cal
 Q}})\;,
\end{eqnarray}
one gets
\begin{equation}
P_i\,|\vec{p},\vec{q}> = (p_i+q_i) |\vec{p},\vec{q}>\;, \label{two2}
\end{equation}
as well, using (\ref{aaa})
\begin{equation}
|\vec{p},\vec{q}> = |\vec{q},\vec{p}>\;. \label{sym}
\end{equation}
We see that the vector (\ref{two}) is characterized by standard
properties of bosonic 2-particle state. We add also that

i) The Abelian addition law (\ref{two2}) and the bosonic symmetry
property (\ref{sym}) as valid as well for arbitrary n-particle state
(\ref{state}) \cite{11},\cite{12},

ii) The complete algebra of $\kappa$-deformed oscillators is
formulated \cite{11},\cite{12} in a way consistent with the
associativity property of the $\circ_{\kappa}$-multiplication.

\section{Discussion.}

In this paper we propose the quantization rules for free
$\kappa$-deformed field theory by introducing the suitably chosen
$\star_{\kappa}$-product of $\kappa$-fields (\ref{newstar1}) and
$\circ_{\kappa}$-multiplication of field oscillators
(\ref{multi0})-(\ref{multi3}). We show that the
$\kappa$-deformed creation and annihilation operators satisfy the
canonical commutation relations with respect to the multiplication
$\circ_{\kappa}$ (see (\ref{aaa0}) and (\ref{aaa})). In such a way
we arrive at the $\kappa$-deformed counterpart of the canonical
field quantization procedure (see (\ref{kappafieldCCR}) and
(\ref{kappafieldCCR1})).

Let us observe that recently the idea of $\kappa$-statistic changing
the momentum dependence of the transposed states was considered in
\cite{10},\cite{nowum2}. In \cite{10} there is constructed
an alternative basis for $\kappa$-deformed Fock space, without considering the respective algebra of $\kappa$-deformed oscillators. An interesting idea was
proposed in \cite{nowum2}. In \cite{11} we started from the
following general form of the $\kappa$-deformed commutativity of the
creation operators
\begin{equation*}
F_{\kappa}(p,q)a_\kappa (f_0(p,q),\vec{f}(p,q)) a_\kappa
\left(g_0(p,q),\vec{g}(p,q)\right)
=\;\;\;\;\;\;\;\;\;\;\;\;\;\;\;\;\;\;\;\;\;\;\;\;\;\;\;\;\;\;\;\;\;\;\;\;\;\;\;\;\;\;\;\;\;\;\;
\end{equation*}
\begin{equation}
\;\;\;\;\;\;\;\;\;\;\;\;\;\;\;\;\;\;\;\;\;\;\;\;\;\;\;\;\;\;\;\;\;\;\;\;\;\;\;\;\;\;\;\;\;\;\;=G_{\kappa}(q,p)a_\kappa
(h_0(p,q),\vec{h}(p,q)) a_\kappa (k_0(p,q),\vec{k}(p,q))
\;.\label{nccr1}
\end{equation}
In \cite{11} we have chose $F_{\kappa} = G_{\kappa} = 1$ and have
selected the functions $f_\mu$, $g_\mu$, $h_\mu$, $k_\mu$ in a
simplest way consistent with $\kappa$-deformed addition law for
fourmomenta and fourmomentum conservation law. In \cite{nowum2}
additionally it was assumed that the relation (\ref{nccr1}) is
covariant under $\kappa$-Poincare boost transformations. It appears that
the $\kappa$-covariant choice of the functions occuring in
(\ref{nccr1}) exist, and satisfy quite involved equations, solved
only perturbatively with respect to powers $\frac{1}{\kappa}$. We
would like to add here that in our approach we transfer the problem
of $\kappa$-covariance of the formalism to the field-theoretic level
(see \cite{12}) where we provided the $\kappa$-covariant commutator
for $\kappa$-deformed free fields.

It should be noted that our $\kappa$-deformed multiplication rules
proposed firstly in \cite{11} are based on nonstandard idea of the
dependence of the threemomentum of one particle on energies of other
particles present in the system. This dependence is obtained  in
such a way that one can introduce the n-particle states which are
symmetric and with classical Abelian addition law for the particle
fourmomenta. One can interprete this energy dependence as a geometric
interference of free field quanta in the trans-Planckian region - in
other words we conjecture that one gets a sort of purely
geometric interaction between particles as a kinematical effect
caused by the space-time noncommutativity \cite{nowicki}.

Finally, it should be pointed out that our considerations in this
note are restricted to the case of free fields only. Our main aim is
to obtain the perturbative framework for the $\kappa$-deformed local
field theory, but this problem is still under investigation.

\end{document}